\documentclass[aps,prd,twocolumn,amsmath,amssymb,amsfonts,nofootinbib,superscriptaddress,altaffilletter]{revtex4-1}
\usepackage{graphicx}
\usepackage{hyperref}
\usepackage{verbatim}
\hypersetup{
bookmarksopen=true
}
\usepackage{amssymb}
\usepackage{amsmath}
\usepackage{color}
\usepackage{supertabular}
\usepackage{dcolumn}

\def\be{\begin{equation}}
\def\ee{\end{equation}}
\def\bea{\begin{eqnarray}}
\def\eea{\end{eqnarray}}

\newcommand{\abs}[1]{\left|#1\right|}

\def\Journal#1#2#3#4{{#1} {\bf #2}, #3 (#4)}

\def\PRL{\em Phys. Rev. Lett.}
\def\PRD{{\em Phys. Rev.} D}

\def\CQG{\em Classical Quantum Gravity} 

\def\APJ{\em  Astrophysical Journal}
\def\PRX{{\em Phys. Rev.} X}

\begin{document}

\title{Impact of signal clusters in wide-band searches for continuous gravitational waves}

\author{ Lorenzo Pierini$^{1,2}$, Pia Astone$^{2}$, Cristiano Palomba$^{2}$, Aidan Nyquist$^{3}$, Simone Dall'Osso $^{2}$, Sabrina D'Antonio$^{4}$, Sergio Frasca$^{2}$, Iuri La Rosa$^{1,2,5}$, Paola Leaci$^{1,2}$, Federico Muciaccia$^{2}$, Ornella J. Piccinni$^{6}$, Luca Rei$^{7}$.  \\
$  $ \\
$^1${Universit\`a di Roma La Sapienza,  Roma, Italy};\\
$^2${INFN, Sezione di Roma, Roma, Italy};\\
$^3${North Park University, Chicago, Illinois, USA};\\
$^4${INFN, Sezione di Roma Tor Vergata, Roma, Italy};\\
$^5${University of Savoia Mont Blanc and  CNRS/IN2P3 Annecy France};\\
$^6${Institut de Física d’Altes Energies (IFAE), Barcelona Institute of Science and Technology, E-08193 Barcelona, Spain};\\
$^7${INFN, Sezione di Genova, Genova, Italy}.\\}

\begin{abstract}
In this paper we present a study of some relevant steps of the hierarchical frequency-Hough (FH) pipeline, used within the LIGO and Virgo Collaborations for wide-parameter space searches of continuous gravitational waves (CWs) emitted, for instance, by spinning neutron stars (NSs). Because of their weak expected amplitudes, CWs have not been still detected so far. These steps, namely the spectral estimation, the {\it peakmap} construction and the procedure to select candidates in the FH plane, are critical as they contribute to determine the final search sensitivity. Here, we are interested in investigating their behavior in the (presently quite) extreme case of signal clusters, due to many and strong CW sources, emitting gravitational waves (GWs) within a small (i.e. $<1$ Hz wide) frequency range. This could happen for some kinds of CW sources detectable by next generation detectors, like LISA, Einstein Telescope and Cosmic Explorer.  Moreover, this possibility has been recently raised even for current Earth-based detectors, in some scenarios of CW emission from ultralight boson clouds around stellar mass black holes (BHs). 
We quantitatively evaluate the robustness of the FH analysis procedure, designed to minimize the loss of single CW signals, under the unusual situation of signal clusters. Results depend mainly on how strong in amplitude and dense in frequency the signals are, and on the range of frequency they cover. We show that indeed a small sensitivity loss may happen in presence of a very high mean signal density affecting a frequency range of the order of one Hertz, while when the signal cluster covers a frequency range of one tenth of Hertz, or less, we may actually have a sensitivity gain.  
Overall, we demonstrate the FH to be robust even in presence of moderate-to-large signal clusters. 

\end{abstract}

\maketitle

\section{Introduction}
Searches for continuous gravitational waves (CWs), emitted for instance by rapidly rotating neutron stars (NSs), either isolated or binary systems (see \cite{ref:lasky,ref:CWreview} for reviews of sources and possible emission mechanisms), are among the highest priorities of the LIGO-Virgo-KAGRA Collaboration ~\cite{ref:WP}. The third scientific run of the LIGO-Virgo advanced detectors (O3) took place from April 1, 2019  to the end of March 2020. CW signals have a duration that is longer than the typical run time of the detectors and contain important information on the nature and dynamics of the source. Presently, no CW source has been detected, due to the weakness of this kind of GW emission, compared to that of compact binaries coalescence \cite{ref:GWTC1,ref:GWTC2,ref:GWTC2.1,ref:GWTC3}.

CW sources we are looking for fall into five broad categories \cite{ref:WP}: (1) nonaccreting known pulsars for which timing data are available; (2) other known or suspected isolated neutron stars, with known sky location but limited or absent timing information; (3) unknown isolated NSs in any direction; (4) accreting stars in known/unknown binary systems; (5) long-transients (with duration of hours/days/months), due for example to a postmerger newborn NS, to r-mode instabilities or to newborn highly magnetized NSs (magnetars) \cite{ref:GW170817PM,ref:LtransO3}. Moreover, recently a growing interest has been focused on boson clouds forming around spinning BHs, as potential CW sources \cite{ref:bosonTEOR}.
The data analysis method used is different for each of these targets. For unknown sources, i.e. when the source parameters (namely, position in the sky, GW frequency, frequency time variation) are not known, fully coherent searches are computationally unfeasible. For this reason, hierarchical procedures (which alternate coherent and incoherent analysis steps) have been set up. Recent results on all-sky CW searches using data of the second and third scientific run of the LIGO and Virgo detectors  (O2 and O3) have been reported in \cite{ref:allskyO2}, \cite{ref:allskyO3a}, \cite{ref:allskyO3}. A short review of the main LIGO-Virgo all-sky CW search methods is also presented in \cite{ref:CWmethods}. 

One of the standard search methods for all-sky searches is the frequency-Hough (FH) pipeline, described in detail in \cite{ref:FH}. The FH pipeline is a hierarchical procedure that aims at identifying the most significant CW candidates in a wide parameter space. A follow-up procedure, described in \cite{ref:allskyO2}, is then applied to these candidates, in order to confirm or reject them. Recently, a GPU implementation of the FH was run to analyze the highest, and computationally demanding, part of the O3 all-sky search, showing a computational gain of more than one order of magnitude \cite{IuriGPU}.
The FH hierarchical procedure strongly reduces the computational cost with respect to a fully coherent search, but at the price of a sensitivity loss. In particular, some thresholds and selection criteria are applied at various stages of the analysis. Any possible signal lost during one of these steps will not be recoverable by later steps. For this reason, each step of the procedure is designed to reduce as much as possible the loss of possible CW signals, which could be hidden by the presence of strong disturbances or even by other superposing signals. 
\\

In this paper we review the procedures for spectral estimation, {\it peakmap} (PM) construction and candidate selection applied on recent LIGO-Virgo data, with the aim of testing their effectiveness when clusters of signals are present. All these steps play a relevant role in determining the signal detectability and may bring to unwanted sensitivity losses if not properly addressed. In particular, by adding simulated signals to O2 data, we demonstrate the robustness of the FH pipeline in presence of signal clusters, quantifying the very qualitative claims reported in \cite{ref:papaetc}.
\\

The paper is organized as follows. In Sec. \ref{sec:fh} we briefly remind the main steps of the FH pipeline. In Sec. \ref{sec:analysis} we discuss the main issue this paper deals with, i.e., the possibility that some steps of the FH analysis are negatively affected by the presence of several concurrent large CW signals confined in a very small frequency range. Section \ref{sec:spectre} is devoted to describe in detail two relevant analysis steps, i.e., spectral estimation and construction of the time-frequency PM, which could be affected by the presence of dense clusters of signals. Sections \ref{sec:efficy} and \ref{sec:effstudy} are the core of the paper and are focused on the detailed studies we have done, by means of software simulated signals added to O2 data, to quantify the effect on the detection efficiency of dense clusters of signals. We end with a discussion on the results in Sec. \ref{sec:conc}. 

\section{The FH search procedure}
\label{sec:fh}
\begin{figure}[htbp]
\includegraphics[width=9cm]{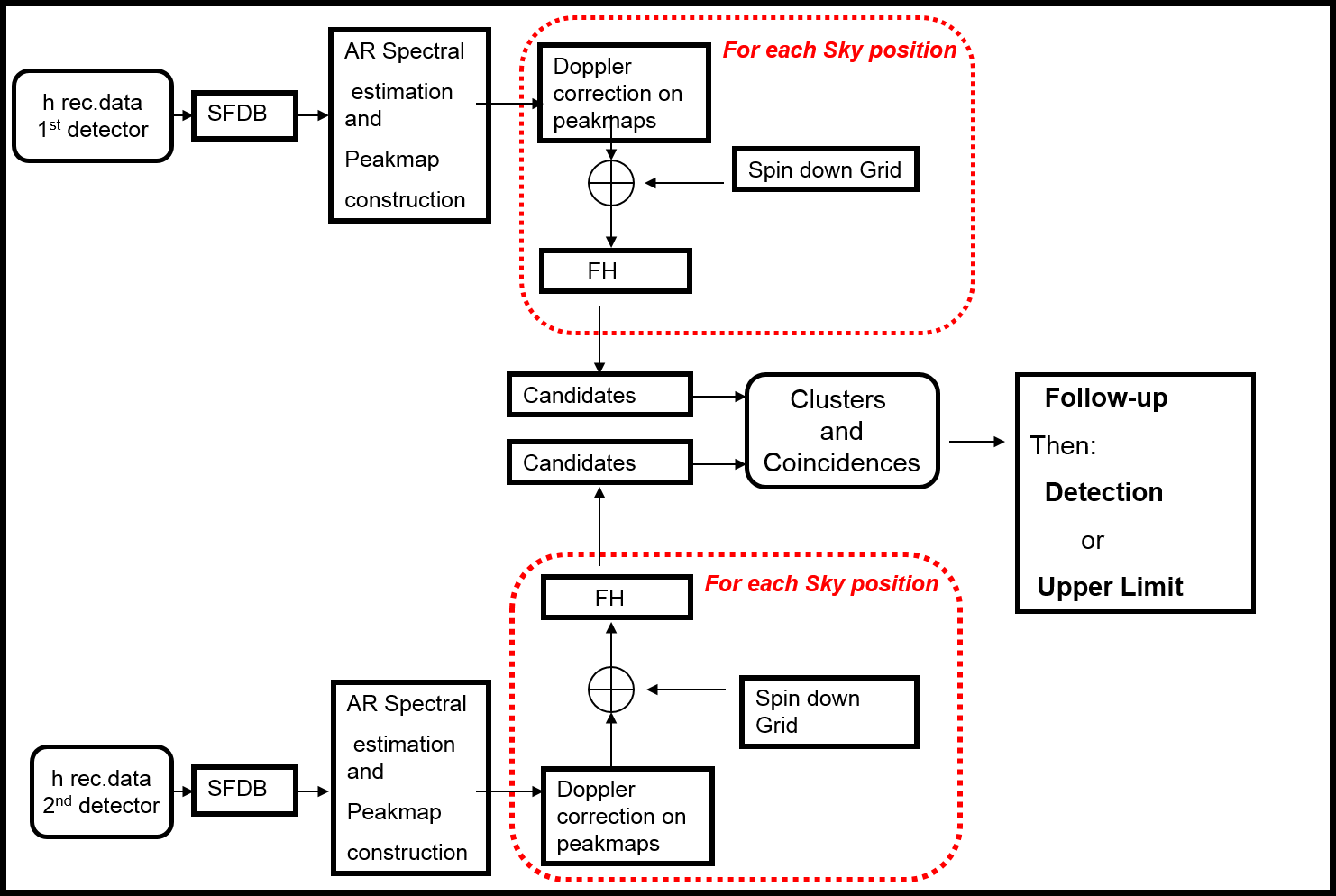}
\caption{Scheme of the hierarchical FH procedure, in the case of data from 2 detectors. See Sec. \ref{sec:fh} for a short description of the steps of the procedure. }
\label{fig:HIER}
\end{figure}

A scheme of the FH pipeline is given in figure \ref{fig:HIER} and is briefly described in the following.
\begin{itemize}
\item{ [\bf A]} Creation of the short fast Fourier transform (FFT) database (SFDB).

Detector calibrated data are split into chunks of duration 8192, 4096, 2048 and 1024 seconds. Each chunk is then Fourier transformed, using the FFT algorithm, obtaining four different sets of frequency domain data covering, respectively, the range [0-128] Hz, [128-512] Hz, [512-1024] Hz, and [1024-2048] Hz. At each FFT chunk are also associated additional data, such as position and velocity of the detector at the chosen reference time.
 There is a work in progress to make this step in a more flexible way, making use of the ``band sampled data" (BSD) \cite{BSD} framework. 

\item{ [\bf B]} Autoregressive (AR) spectral estimation. 

An AR algorithm, described in \cite{ref:pia_sfdb}, is used to smooth the noise spectral density of each chunk of data in such a way to preserve possible CW signals, which in a single FFT would be confined within one single frequency bin, and at the same time to be able to follow slower spectral variations. The filtering is usually done backwards through the frequency bins, that is from higher to smaller frequencies, to better adapt recursion to the noise behavior of the detectors, which shows much higher and rapidly varying values in the low frequency band, i.e., below $\approx$ 30 Hz.

\item{ [\bf C]} PM creation.

For each FFT, local maxima above a given threshold, called peaks, are selected in the ratio $\cal{R}$ among the periodogram (square modulus of the FFT) and the AR spectrum estimation. The PM is the collection of all the peaks, each identified by a value of the frequency and a time (the middle time of the corresponding FFT), both measured at the detector. AR spectral estimation and PM construction have been used since many years for these analyses and their mathematics (and statistics) is described in \cite{ref:pia_sfdb}.

\item{ [\bf D]} FH transform.

For each sky position, the peaks in the PM are shifted in frequency to remove the Doppler effect for that position at the peak time. The shifted peaks are fed to the Hough transform, which maps the time frequency peaks to the source frequency and spin-down plane. 

\item{ [\bf E]} Candidate selection and coincidence analysis.

Candidates are selected in the FH plane by finding, for each sky position, the points with highest number count in each frequency and spin-down subinterval.
Candidates from the analysis of a given data set are clustered together, and
cross-checked against candidates found in the analysis of another data set (of the same detector or of a different detector) by means of coincidences in the signal parameter space. 

\item{ [\bf F]} Candidates follow-up and verification.

A deeper analysis is done on those candidates which survive the coincidence step, in order to check their astrophysical origin or demonstrate they are compatible with noise or due to detector artifacts. The standard follow-up approach consists in rerunning the search starting from longer duration FFTs, in a limited portion of the parameter space around each candidate.
In case of no detection, upper limits (UL) on the signal strain amplitude are computed as a function of the frequency.
\end{itemize}

\section{Boson clouds constraints from FH search on O2 data}
\label{sec:analysis}
Recent results obtained for the O2 all-sky CW search \cite{ref:UL} have also been used to derive constraints for a class of CWs emitters different from spinning NSs, namely ``clouds'' of ultralight bosons around black holes (BHs). Such  clouds are expected to emit CWs at a frequency which depends at first order on the boson mass and at second order on the product of the boson mass and BH mass (as explained in \cite{ref:UL} and references therein).
Even if more specific methods have been developed to search for this kind of signals, see for example \cite{ref:nostroBOS,ref:bosonO3}, it is possible to map the results of a standard all-sky CW search into exclusions limits in a plane defined by the mass of the scalar boson field and the mass of the BH, as discussed in \cite{ref:UL}.
The basic assumption is that all the steps of the hierarchical FH procedure, especially designed and tested for the weak and rare CW signals emittted by isolated NSs, are compatible with the characteristics of the boson cloud population. In particular, we need to verify if  the sensitivity is degraded when the emitted signals are clustered in frequency. Such scenario could arise under optimistic assumptions about the number of stellar mass BHs present in our Galaxy \cite{ref:BHpop1,ref:BHpop2}: if we assume that the boson clouds formation mechanism affects a significant fraction of galactic BHs, we could have a large number of concurrent CW emitters. These signals would have proper frequency centered at a value corresponding to the boson mass (which reasonably takes just one value), with a spread due to the BH mass distribution. 
If many signals are concentrated in a small frequency range and the signals are ``strong'' enough\footnote{There are presently estimations which give quite different results, depending on the basic assumptions. See for example \cite{ref:BHpop1,ref:BHpop2}.}, it could happen that the stronger signals hide the weaker ones or that there is some degradation of sensitivity due to the fact that signals mix,  making it difficult to separately identify them.
Such possible complication might happen especially in future GW detectors, like Cosmic Explorer \cite{ref:futureCE} and the Einstein Telescope \cite{ref:futureET}, which will be the first pan-European ground-based GW antenna \cite{ref:adele}. These 3G ground-based detectors will achieve one order of magnitude better sensitivities with respect to the existing detectors. On the other side, the space-based detector LISA \cite{ref:futureL} will give us access to the mHz frequency band.

In a recent publication \cite{ref:papaetc}, the impact of signal clusters on the spectral estimation used in the FH approach has been very \textit{qualitatively} discussed. Specifically, plots suggesting the AR procedure cancels signal peaks have been shown, when the signals are concentrated in such a way to produce bumps in the detector noise, thus reducing the search sensitivity. We will demonstrate these conclusions are wrong in most cases.

We quantify these qualitative predictions, studying how spectral estimation, PM construction and candidate selection behave in this situation. In fact, these are the steps of the procedure that might suffer from problems related to the presence of many strong signals. We inject fake CW signals into O2 data to mimic this effect and to see if, and at which level of density and strength of the signals, we might need to apply modifications to the procedure. For simplicity, this study is  done using one week of data, which is enough to evaluate possible problems and sensitivity losses. In fact, by analyzing a longer data set the discriminatory power of the Doppler modulation, see Sec. \ref{sec:efficy}, would be even stronger making our robustness tests conservative. 

\section{Spectral AR estimation and PM construction }
\label{sec:spectre}
The first important step with an impact on the sensitivity of the hierarchical FH procedure is the estimation of the average power spectral density. The procedure is described in details in \cite{ref:pia_sfdb} and we briefly remind the reader here its basic aspects, and give some examples using O2 data. A nearly monochromatic wave with enough high amplitude will produce a delta spike above the noise floor in the periodogram. That peak will be included into the PM only if its power is greater than the corresponding point in the AR estimation by a fixed factor. So, if the AR estimator follow that peak, its chances to be selected in the PM decrease. So, a good spectral estimator for CW searches should have the following properties: 
\begin{itemize}
    \item If narrow peaks in the frequency domain are present, the estimator should not be affected by that peaks. This should be as much as possible independent on the signal-to-noise ratio (SNR) of the peak.
    \item If the noise level varies, either slowly or rapidly, the estimator should be able to follow the noise variations.
\end{itemize}

\begin{figure*}  
\includegraphics[width=8.5cm,height=6cm]{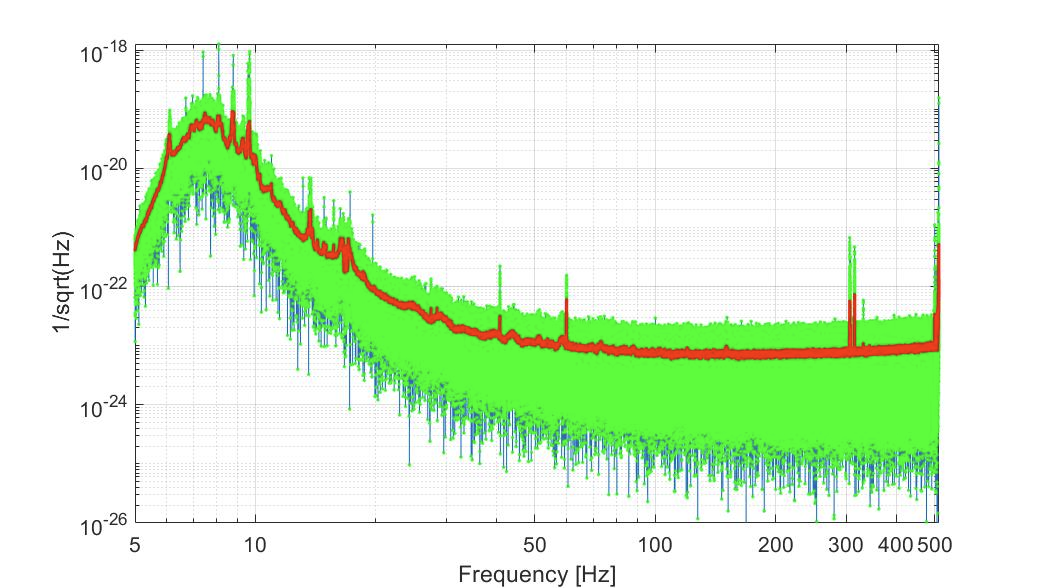}
\includegraphics[width=9cm,height=6cm]{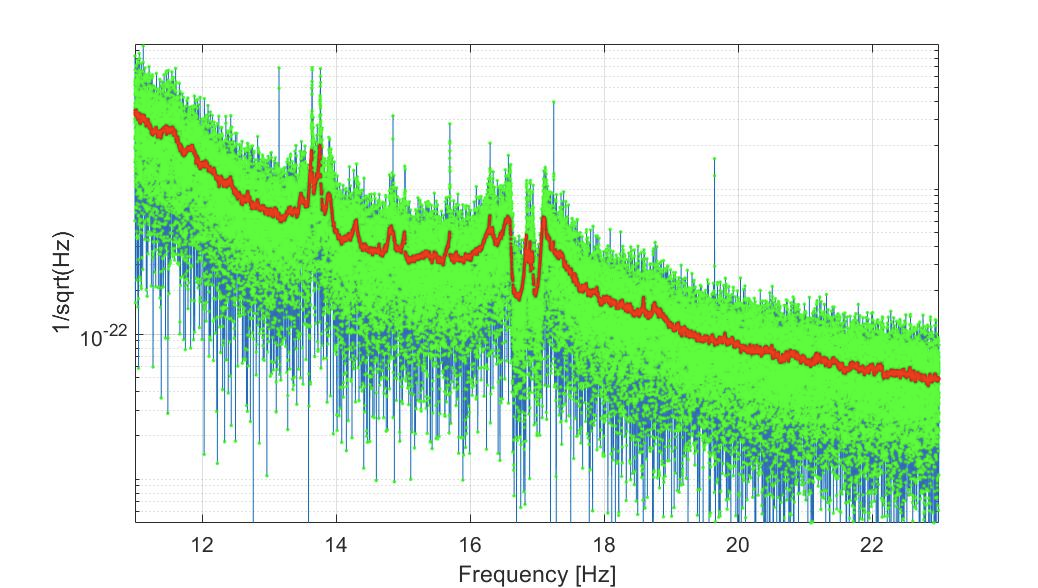}
\caption{An example of one FFT (green dots, blue lines), of duration 8192 s, done using LIGO Livingston O2 data \cite{ref:GWOSC}, and the corresponding AR amplitude spectral density estimation (red). Here, for comparison, the FFT has been normalized to represent the amplitude spectral density and we show the absolute value. The left plot covers the band [5-512] Hz and the second plot is a zoom in the frequency region from 11 Hz up to 23 Hz. It is possible to appreciate that the AR estimation follows sharp changes on noise floor, like those around 14 Hz and 17 Hz. On the other side, narrow peaks like those at 13.15 Hz, 14.85 Hz, 17.25 Hz and 19.65 Hz do not appear in the AR estimation, as required.}
\label{fig:AR1}
\end{figure*}

Let $x_i$ be the data samples of the FFT. The amplitude spectral density is estimated from an AR estimation, as shown in the equations below:
$$y_i=x_i+w \cdot y_{i-1}$$
$$w=e^{-\delta \nu/\tau_f}$$
where $y_i$ are the samples of the not normalized AR mean, 
obtained using $w$ as weight, with $\delta \nu$ being the FFT resolution and $\tau_f$ the memory of the AR mean (with dimensions of a frequency). 
The values of the normalized AR estimations are given by
$$\mu_i=\frac{y_i}{Z_i},$$ where the normalization constant is $Z_i=1+w \cdot Z_{i-1}$.
Besides this, in order to have a ``clean" estimator, that is not affected by spectral peaks, we define a threshold 
$V_{max}$ and an age $A_{max}$.
While $r=\frac{x_i}{\mu_{i-1}}$ is lower than (or equal to)  the threshold, the new datum $x_i$ is used to evaluate the actual mean and
the age of the estimator is set to zero (expressed in number of samples). \\
When $r=\frac{x_i}{\mu_{i-1}}$  is larger than the threshold,  the new datum is not used to evaluate the actual mean 
and the age of the estimator is incremented by 1 bin (that is by a frequency equal to the resolution). This eliminates or at least reduces the effect of peaks from the estimation.\\
If the estimator becomes too old, i.e., if the age becomes greater than the maximum age we have set, we deduce that we are not in presence of a peak (and thus of a possible signal), but only that the noise characteristics have changed.
We thus go back by a number of samples $n=A$, and begin a new evaluation
of the mean, restarting from zero at the sample $i-A$. This is needed to deal with all those situations when the noise is highly non-stationary.
Figure \ref{fig:AR1} shows an example, where the AR estimation is plotted together with the absolute value of the corresponding FFT. From the plots on the left and from the zoom on the right it is possible to note how, when narrow peaks are present in the FFT, they do not appear, or appear significantly reduced, in the AR estimation, and so are not (or only slightly) suppressed in the PM. 

More examples are given in Sec.~\ref{sec:efficy}, where we show the results after adding fake CW signals to the O2 data.

\section{Signal clusters: preliminary considerations}

\label{sec:efficy}
As anticipated, we want to study the robustness of our method. In the following, we will focus on the AR spectral estimation and PM characterization steps, showing how they behave in presence of many CWs.
\subsection{Qualitative effect on AR estimation}
\begin{figure*}
\includegraphics[width=8cm,height=5cm]{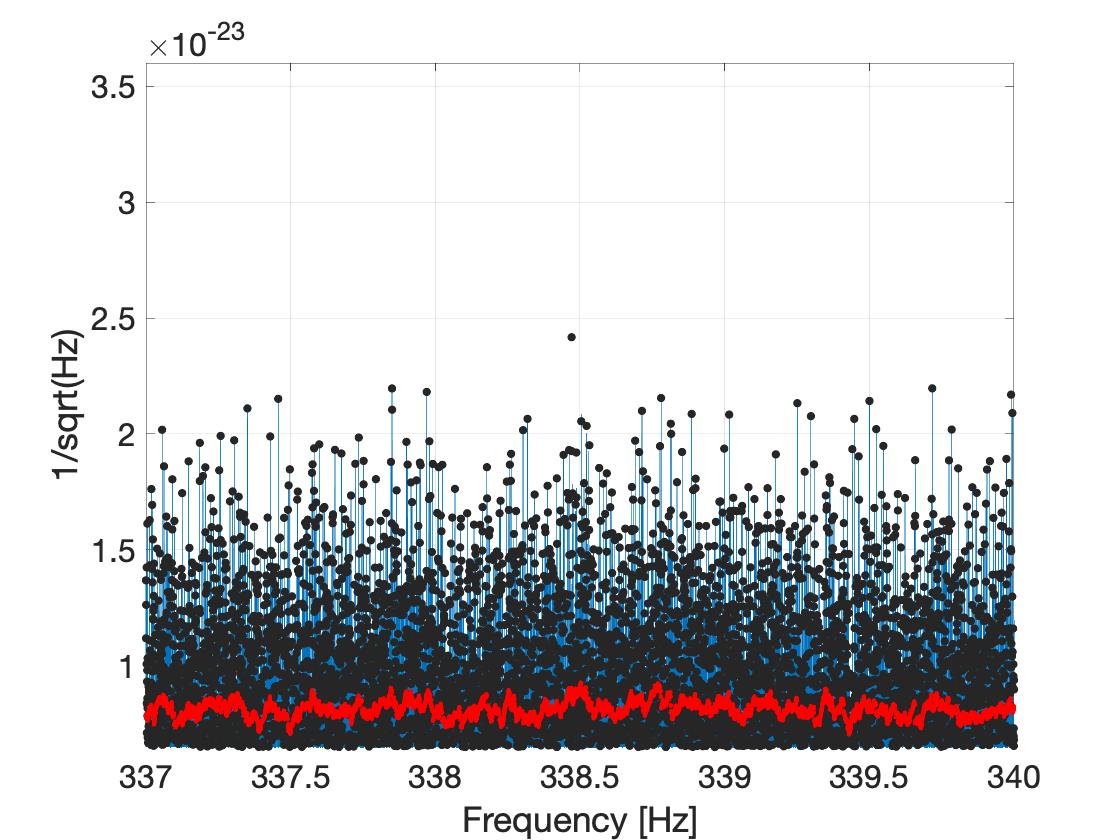}
\includegraphics[width=8cm,height=5cm]{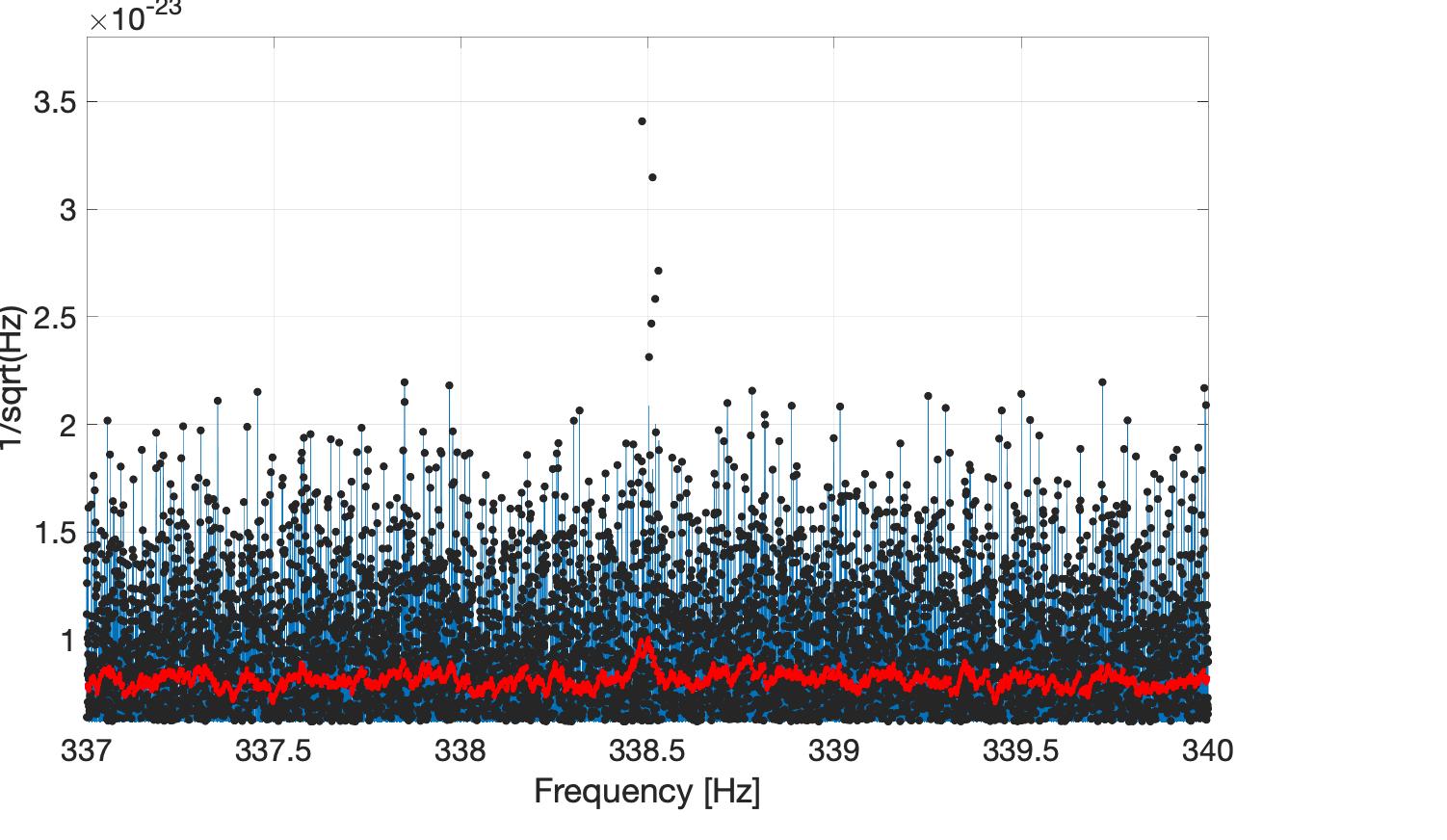}
\includegraphics[width=8cm,height=5cm]{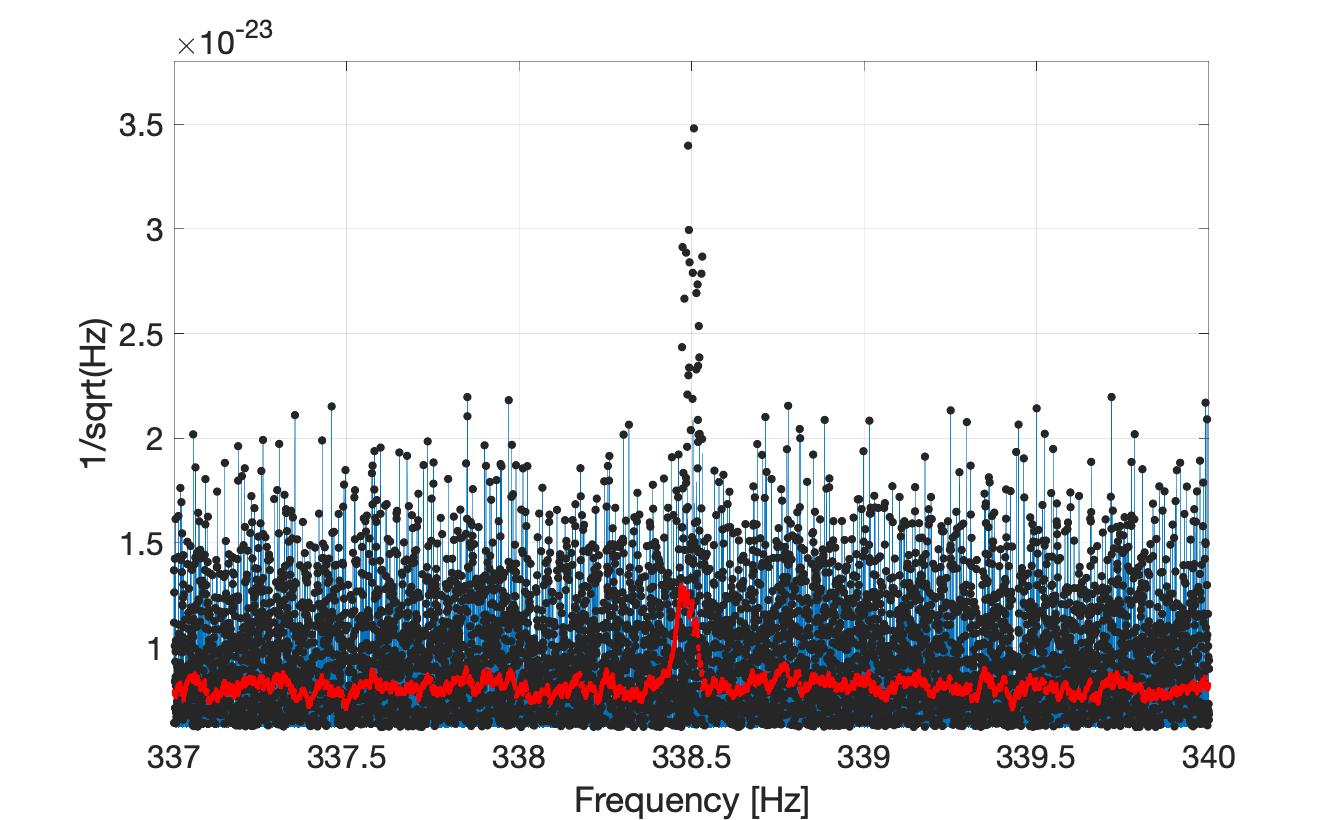}
\includegraphics[width=8cm,height=5cm]{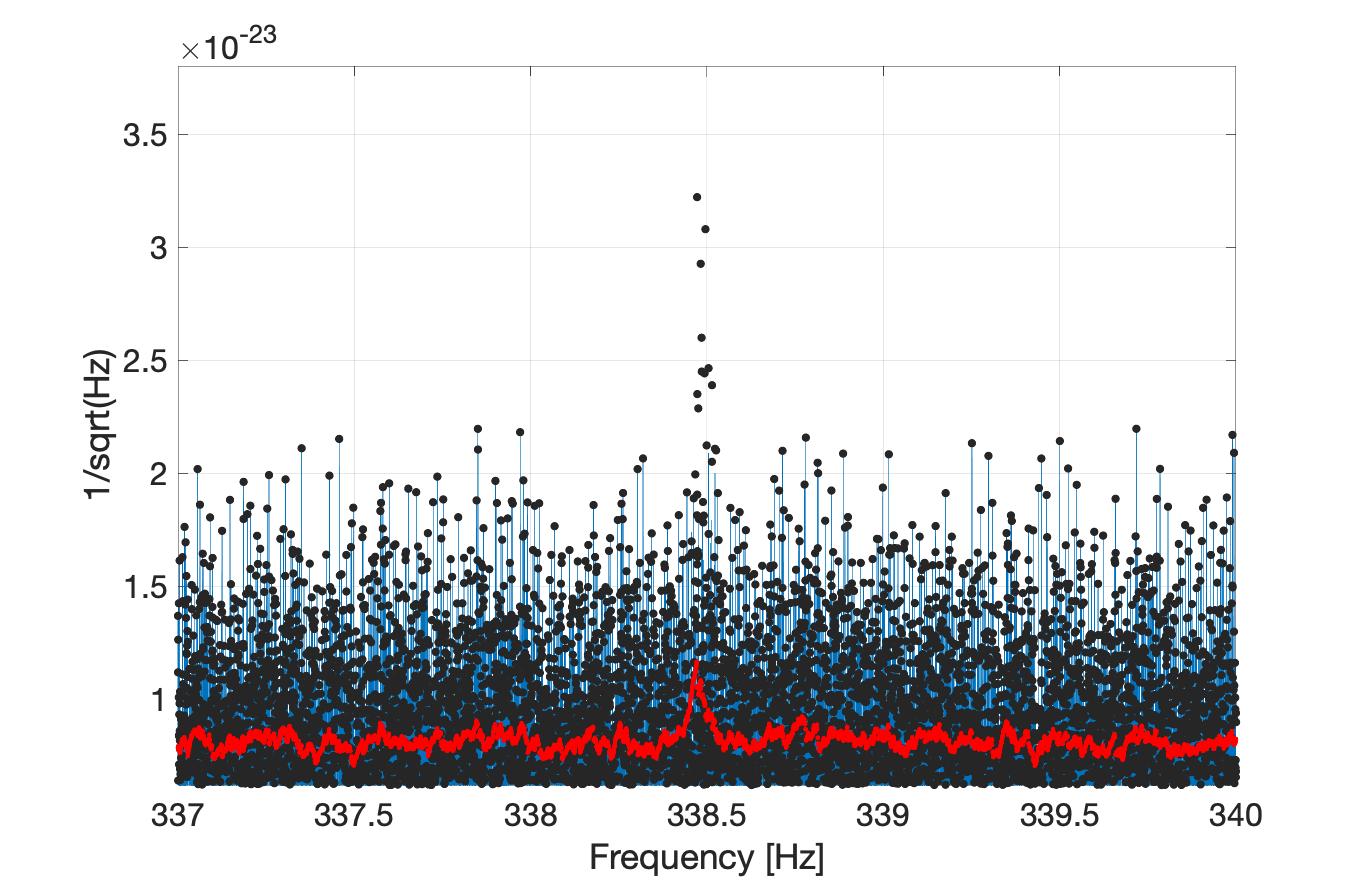}

\caption{Plots of the 4 different injections listed in Table \ref{table:1} made on one FFT calculated using 4096 s of LIGO Livingston O2 data \cite{ref:GWOSC}. Top left, top right, bottom left, bottom right show the configurations respectively of label 1, 2, 3 and 4.Blue lines and black dots represent the absolute value of one normalized FFT, in a 3 Hz band, around 338.5 Hz. The red line is the corresponding AR amplitude spectral density estimation. The effect of the signal peaks in the AR estimation is negligible in the first two cases, very small in the last case, while it reduces the contribution of some of the strongest peaks in the third case (50 strong signals, with source frequency $f_0$ within 5 bins).}
\label{fig:inj}
\end{figure*}

In what follows, we have generated and injected several CW signals, with randomly chosen sky localization, nearby source frequencies and, for simplicity, without spin-down. 
The range of signal amplitudes we have considered is $2 \times 10^{-25} - 10^{-24}$, which is above the ULs found in O2 \cite{ref:UL} at the chosen frequencies, and corresponding to the range predicted in \cite{ref:papaetc} for the detectable galactic boson clouds/BHs population. Please note that the signal amplitude is well above the noise level: if the signals were weaker, the cumulative effect on AR estimation and on the overall procedure would be even lower.\\
Figure \ref{fig:inj} shows what happens to the AR estimation inside a single FFT, in 4 different injection configurations. The FFTs are built using 4096 seconds of LIGO Livingston O2 data \cite{ref:GWOSC}, so their frequency resolution (as well as for the AR estimation) is $\delta\nu=1/4096\rm s\simeq 2.44\cdot 10^{-4}~\rm Hz$. All added signals have their proper frequencies $f_0$ within few frequency bins, but because of their randomly distributed sky position they are further spread by Doppler modulation. In details, the configurations are in Table \ref{table:1}.
\begin{table}[h!]
\centering
 \begin{tabular}{|c c c c|} 
 \hline
 label & n.sig & $f_0$ range [Hz] & $h_0$ \\ [0.5ex] 
 \hline
 1 & 10 & $\left[338.5,~333.5+\delta\nu\right]$ & $1 \times 10^{-24}$ \\ [0.5ex]
 2 & 20 & $\left[338.5,~333.5+2\delta\nu\right]$ & $1 \times 10^{-24}$ \\[0.5ex]
 3 & 50 & $\left[338.5,~333.5+5\delta\nu\right]$ & $1 \times 10^{-24}$ \\[0.5ex]
 4 & 50 & $\left[338.5,~333.5+5\delta\nu\right]$ & $6 \times 10^{-25}$ \\[0.5ex] 
 \hline
 \end{tabular}
\caption{Parameters used to simulate CWs in different density configurations, labeled from 1 to 4.}
\label{table:1}
\end{table}

By looking at the AR estimation (red curve in Fig. \ref{fig:inj}), it is clear that generally it is negligibly affected by the presence of an ensemble of signals at low densities: the only case when it is affected is when we add 50 strong signals with $f_0$ within 5 bins, which means within 1.23 mHz. In this case, the ratio $\cal{R}$ obtained on the central peaks is reduced by $\sim$30\%.
This fact obviously motivates a further investigation. However, as it will be shown in the next subsection, this amplitude reduction does not automatically affect the sensitivity of the search, as 
\begin{itemize}
    \item it is not affecting all the peaks and, most of all, not always the same peaks when their time evolution is considered;
    \item the amplitude of the peaks is never used in the analysis, so what matters is only the limit posed by the threshold used in the PM creation.
\end{itemize}
These aspects are discussed in more detail in the following subsection.

\subsection{Peakmap characterization}
\begin{figure*}
\includegraphics[width=8.5cm,height=6cm]{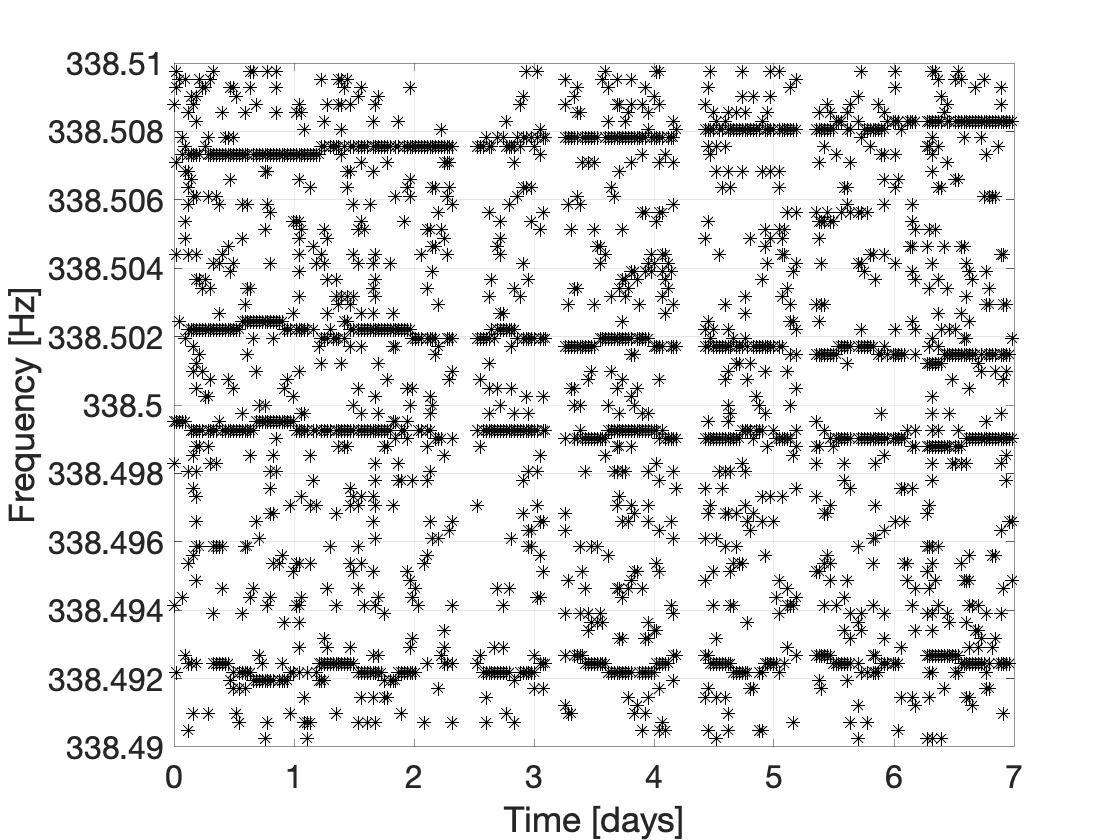}
\includegraphics[width=8.5cm,height=6cm]{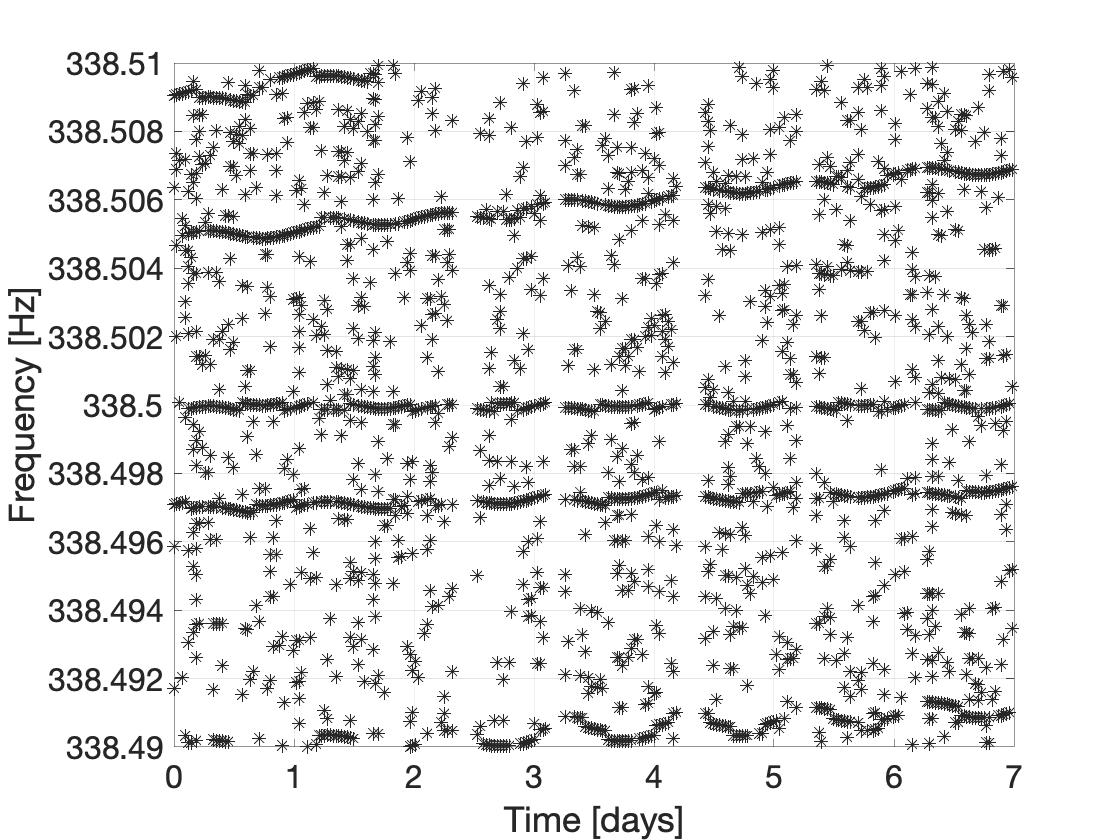}
\includegraphics[width=8.5cm,height=6cm]{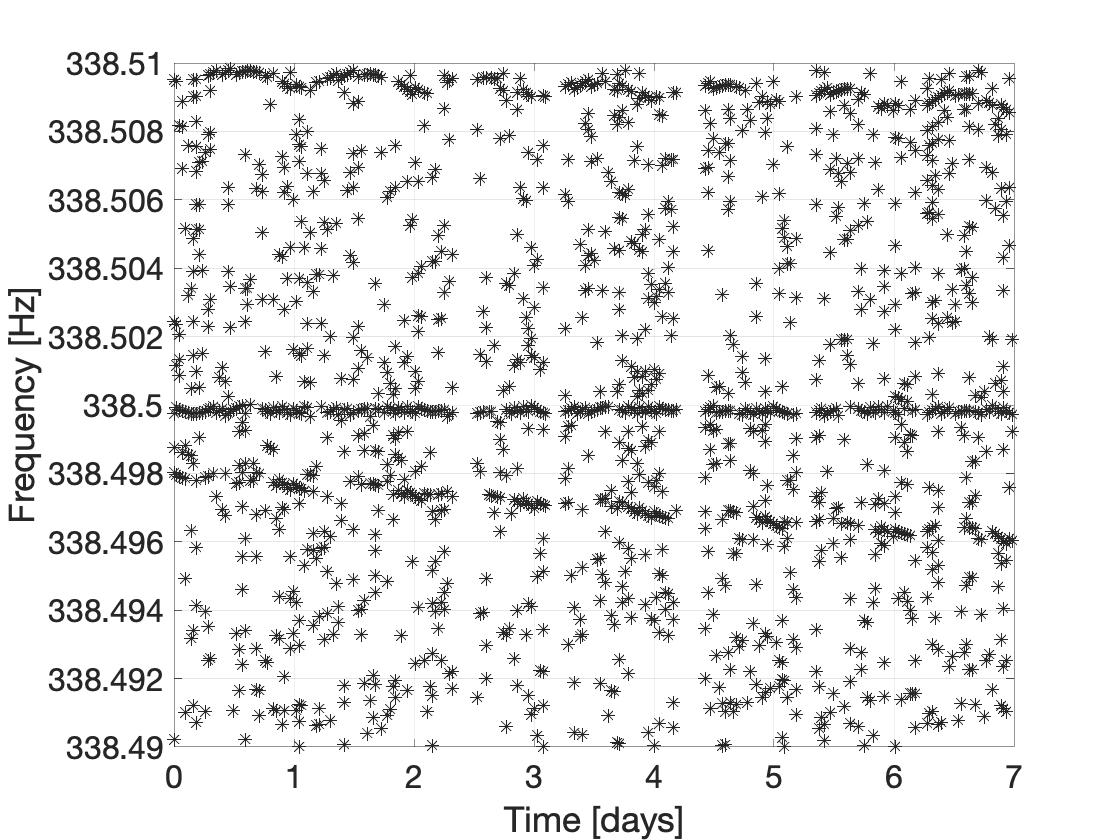}
\includegraphics[width=8.5cm,height=6cm]{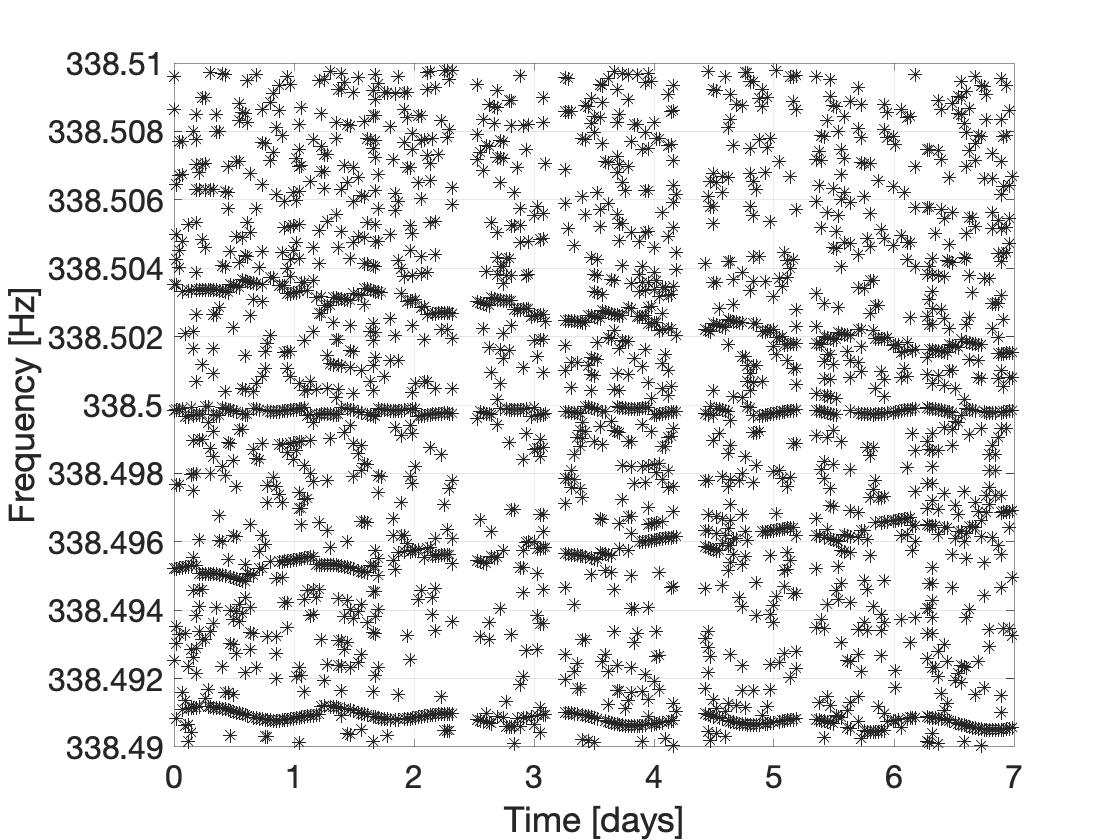}
\caption{The top left plot shows a zoom on a PM of LIGO Livingston O2 data \cite{ref:GWOSC}, where different fake CW signals are added. Those signals have all the same proper frequency, $f_0=338.5$ Hz, but different sky positions. They are strong enough to produce patterns visible by eye: each one follows a different time-frequency trajectory, having different Doppler modulations. In the other plots (top right, bottom left and bottom right) the Doppler correction is applied respectively for one of the signals sky direction. When the Doppler correction matches the sky localization of a signal source, its corresponding trajectory becomes a straight line confined on $f_0$. On the other side, the other signals trajectories get further distortion.}
\label{fig:PM}
\end{figure*}

As anticipated, it is important not to lose possible signal peaks at this level, as they cannot be recovered after. We also notice that in order to build the PM, first the ratio $\cal{R}$ of the periodogram (square modulus of the FFT) to the AR power spectrum estimation is computed. Then, local maxima of $\cal{R}$ above a given threshold are selected. In fact the actual value of $\cal{R}$ for the selected peaks is not used in the analysis. This choice allows us to reduce the impact of large noise spectral disturbances. 
\\A very relevant point is the role of Doppler correction. This is done by properly shifting the peaks frequency in the input PM, for each sky position for which we are running the search. As shown in Figure \ref{fig:PM}, even in the case of multiple signals belonging to the same frequency bin at a given time, the Doppler correction,  which is different for each sky position, permits to clearly distinguish the various signals. This is a very powerful feature as, moreover, it enhances the SNR of CW signals with respect to noise lines. The first plot in Figure \ref{fig:PM} shows a zoom of a PM with 10 injected signals. The signals have been generated with the same intrinsic frequency $f_0$, that is within the same frequency bin, but with randomly chosen sky positions, so that their frequencies at the detector appear separated, due to the Doppler effect which introduces a maximum frequency shift of 
\begin{equation}
\Delta f_{dop}\simeq 2f_0\abs{\frac{\vec{v}\cdot\hat{n}}{c}}\simeq 2f_0\times 10^{-4},
\label{eq:deltafdopp}
\end{equation}
where $\vec{v}$ is the detector velocity vector and $\hat{n}$ is the versor identifying the source position in the sky.
When the Doppler effect is removed correctly for each of these signals, we can see the injected signal as a sequence of peaks along a straight line at the right $f_0$ frequency. This is shown from the second to the fourth plots of Figure \ref{fig:PM}. It is possible to note here that there is only one straight line, which corresponds to the properly Doppler corrected peaks. 
In conclusion, these results show that, even if signals are clustered in frequency, the presence of the Doppler modulation gives a powerful tool to distinguish them. Thus, in all analyses where candidates are selected separately for each sky position, as in the case of the FH pipeline, the presence of signal clusters does not blind the search. Clearly, the resolving power is limited by the resolution of the sky grid used, which depends on the frequency resolution. The candidate separation effect is clearly stronger as the observing time improves, and it is maximum for one year of data, when the Doppler annual modulation completes one cycle.
\section{Efficiency study: Frequency-Hough can handle multiple signals}
\label{sec:effstudy}
\begin{figure*}
\includegraphics[width=10cm,height=6cm]{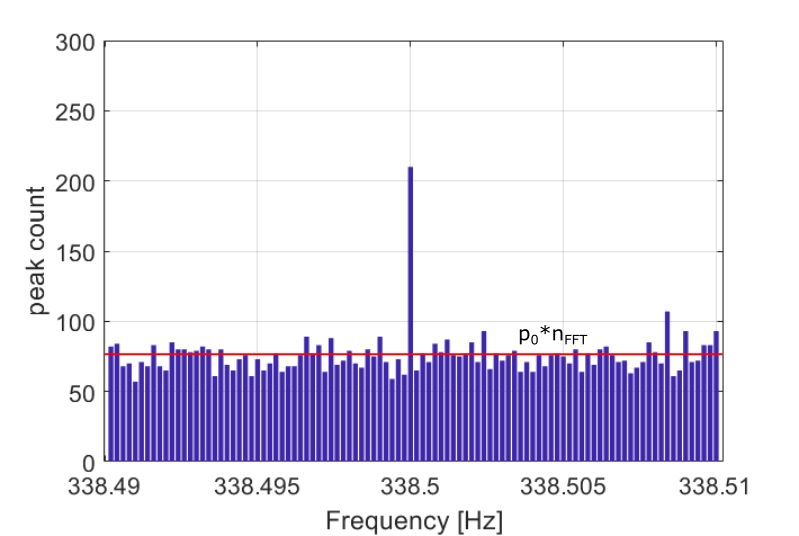} 
\caption{Example of efficiency calculation from full PM after Doppler correction. A histogram of the peaks is built using the frequency resolution as bin width. Here an excess of counts is clearly visible at the frequency $338.5$ Hz, where a signal has been injected into $n_{FFT}=1000$ discrete FFTs. The red line represents the theoretically calculated expected noise counts, $E[\bar{n}_{noise}]=p_0\cdot n_{FFT}$.}
\label{fig:hist}
\end{figure*}

This section is devoted to properly quantify the effect of multiple signals and their impact on the pipeline performances. In order to do this, we need to define an observable. We consider signals that, after the Doppler correction are monochromatic, meaning they are represented in the PM by a straight line at the constant emission frequency $f_0$. So, after the proper corrections, we construct an histogram of the number of peaks as a function of the frequency, using the same frequency resolution of the search. We then expect to find an excess of counts in the histogram bin associated to the source frequency $f_0$. We define the efficiency as
\begin{equation}
\eta =\frac{n_{sig}-\bar{n}_{noise}}{n_{FFT}},
\end{equation}
where $n_{sig}$ is the histogram number count at the frequency $f_0$, $\bar{n}_{noise}$ is the average number of counts at all the other frequencies except for $f_0$, and $n_{FFT}$ is the number of FFTs used to construct the PM.  Figure \ref{fig:hist} shows an example of efficiency versus frequency, having injected a signal at $f_0=338.5$ Hz, constructed the PM and removed the Doppler effect.
We need to underline that the detection efficiency $\eta$ defined here  does not determine the full sensitivity of the FH procedure, even if the two concepts are strictly related. The expected value of $\eta$  is the difference between the probability of selecting a peak when a signal with normalized spectral amplitude $\lambda$ is present, namely $p_{\lambda}$, and the probability of selecting a peak when only Gaussian noise is present, that is $p_0$~\cite{ref:FH}:
\begin{equation}
E[\eta]=p_{\lambda}-p_0.
\end{equation}
In the following, we will estimate $\eta$ in different situations. We use one week of data\footnote{We use only one week of data in order to limit the computational cost of the study, as the goal of the paper is a comparison of different situations on the same data set}, selected from the LIGO Livingston O2 run \cite{ref:GWOSC}, starting from GPS=1186606177. 
The efficiency is computed as a function of the density of signals in given frequency intervals. We call $\Delta f_0$ the interval in which the source-frame frequencies are randomly generated. Because of the random sky localization of the sources, the detector-frame observed frequency will be unevenly distributed into a frequency range given by
\begin{equation}
\Delta f=\Delta f_0\pm \Delta f_{dop}.
\label{eq:deltaf}
\end{equation}
where $\Delta f_{dop}$ is the spread of the signal frequency at the detector frame due to the Doppler effect, given by Eq. \ref{eq:deltafdopp}.
In this way we can establish how many signals to inject in a given frequency interval by tuning $f_0$ and $\Delta f_0$. The number of frequency bins covered by the signals is
\begin{equation}
n_\mathrm{bin}=\frac{\Delta f}{\delta\nu}
\end{equation}
and we can refer to the resulting mean density of signals-per-frequency-bin as
\begin{equation}
\rho_{sig}=\frac{N_\mathrm{sig}}{n_\mathrm{bin}}.
\label{eq:sigdensity}
\end{equation}
In the following subsections we will quantify the results in terms of estimates of detection efficiency, first adding signals with the same amplitude and then with uniformly distributed amplitudes.

\subsection{Constant amplitude signals}
\label{subsec:constamp}
\begin{figure*}
\includegraphics[width=8.5cm,height=6.5cm]{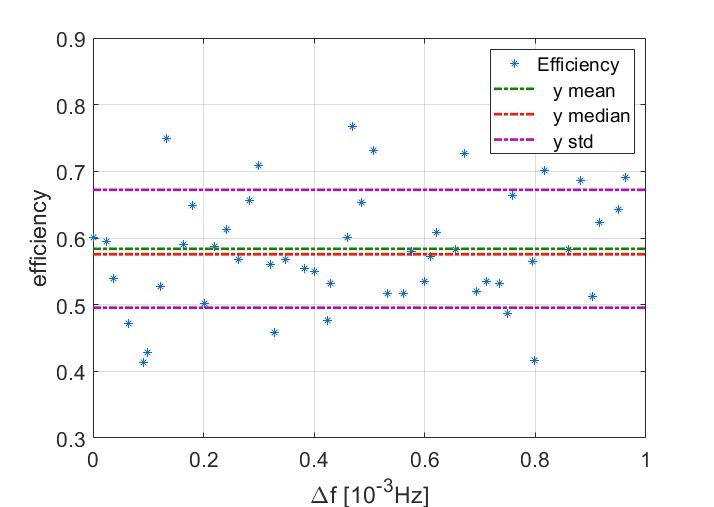}
\includegraphics[width=8.5cm,height=6.5cm]{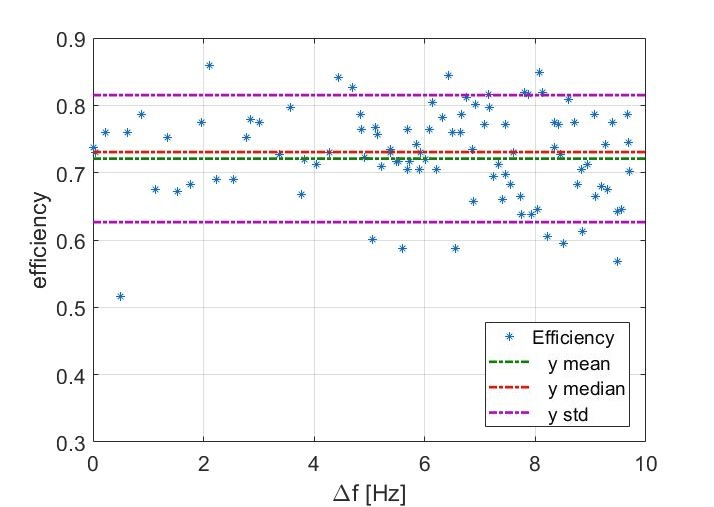}
\caption{Efficiency $\eta$  in the two cases of: 50 injected signals with $h_0=10^{-24}$ distributed in $\Delta f_0=5\delta\nu\sim1.2$ mHz - left plot - and widely spaced (over 10 Hz) - right plot. Mean, median and standard deviation of the efficiency are also shown.}
\label{fig:inj50}
\end{figure*}

As a first case study, we consider a rather extreme scenario. In what follows 50 signals, all with the same very large amplitude, are simulated with proper frequency in a band respectively $5\delta\nu$ and 10 Hz wide. The injection parameters are listed in Table \ref{table:constamp}. Figure \ref{fig:inj50} shows all the obtained efficiencies in both cases, together with mean values, median and standard deviation of the results. Estimations of the overall efficiency, also with the signal-per-bin densities, are reported in Table \ref{table:constamp}.
\begin{table}[h!]
\centering
 \begin{tabular}{|c c c c c|} 
 \hline
 label & $\Delta f_0$ & $h_0$ & $\rho_{sig}$ & $E\left[\eta\right]\pm\sigma_\eta$ \\ [0.5ex] 
 \hline
 $5\delta\nu$ & $\left[338.5 , 338.5 +5\delta\nu\right]$ & $1 \times 10^{-24}$ & $\sim0.15$ & $0.58\pm0.09$ \\ [0.5ex]
 10Hz & $\left[333.5 , 343.5\right]$ & $1 \times 10^{-24}$ & $\sim10^{-3}$ & $0.72\pm0.09$ \\[0.5ex]
 \hline
 \end{tabular}
 \caption{Parameters used to simulate CWs in two different density configurations, labeled as $5\delta\nu$ and 10Hz, the obtained signal densities $\rho_{sig}$ and efficiency estimations.}
 \label{table:constamp}
\end{table}

The mean efficiency loss (for the case of nearby signals with respect to the case of widely spaced signals) is $\sim 19\%$. In other words, in that extreme situation we recover in average the $81\%$ of pixels per signal with respect to the widely spaced one. In Sec. \ref{sec:simBC} much higher densities than $\rho_{sig}\sim0.15$ will be explored, but very different efficiency losses will be obtained. The reason is that the constant signal amplitudes well above the noise floor strongly impact the AR estimation. Anyway, we note that the presence of only strong signals seems to be so unrealistic in the context of CWs from boson clouds. We could have a similar scenario in future signal-dominated detectors, like LISA. In the following, we simulate signals with randomly distributed amplitudes.

\subsection{Uniform amplitude signals with low densities}
\label{subsec:lowdens}
In this subsection we present the results of multiple injections in a more realistic regime in which signal amplitudes are not constant, as in the previous case, but uniformly distributed. In order to get a complete understanding, we cover the whole frequency region where the detectors have the best sensitivity, [70 - 512] Hz. The signals have been simulated with source frequencies around 12 different values in that interval and with source positions randomly chosen. For each one of that 12 central frequencies, 10 signals have been injected with proper frequencies distributed in six different ranges $\Delta f_0$. The choice is done in order to search for evidence of a decreasing efficiency when the signals are concentrated in a smaller region. All the chosen parameters are reported below:
\begin{table}[h!]
\raggedright
    \begin{tabular}{c}
     $h_0\in\left[2\times 10^{-25} , 1\times 10^{-24}\right]$ \\ [1ex]
    \end{tabular}
    \begin{tabular}{c c c c c c c c}
        $f_0$ $\in$ & \{70.5; & 105.5; & 120.5; & 170.5; & 205.5; & 220.5; & \\
        & 270.5; & 330.5; & 380.5; & 420.5; & 442.5; & 492.5\} & Hz \\ [1ex] 
        $\Delta f_0\in$ & \{$\delta\nu$; & $10\delta\nu$; & $20\delta\nu$; & $50\delta\nu$; & $100\delta\nu$; & 10 Hz\} &
    \end{tabular}
\end{table}\\
Thus, the signal densities $\rho_{\mathrm{sig}}$ explored in this section are in the range [0.001 - 0.1]. We can refer to this part as a "low density" regime.

\begin{figure*}
\includegraphics[width=17cm,height=10cm]{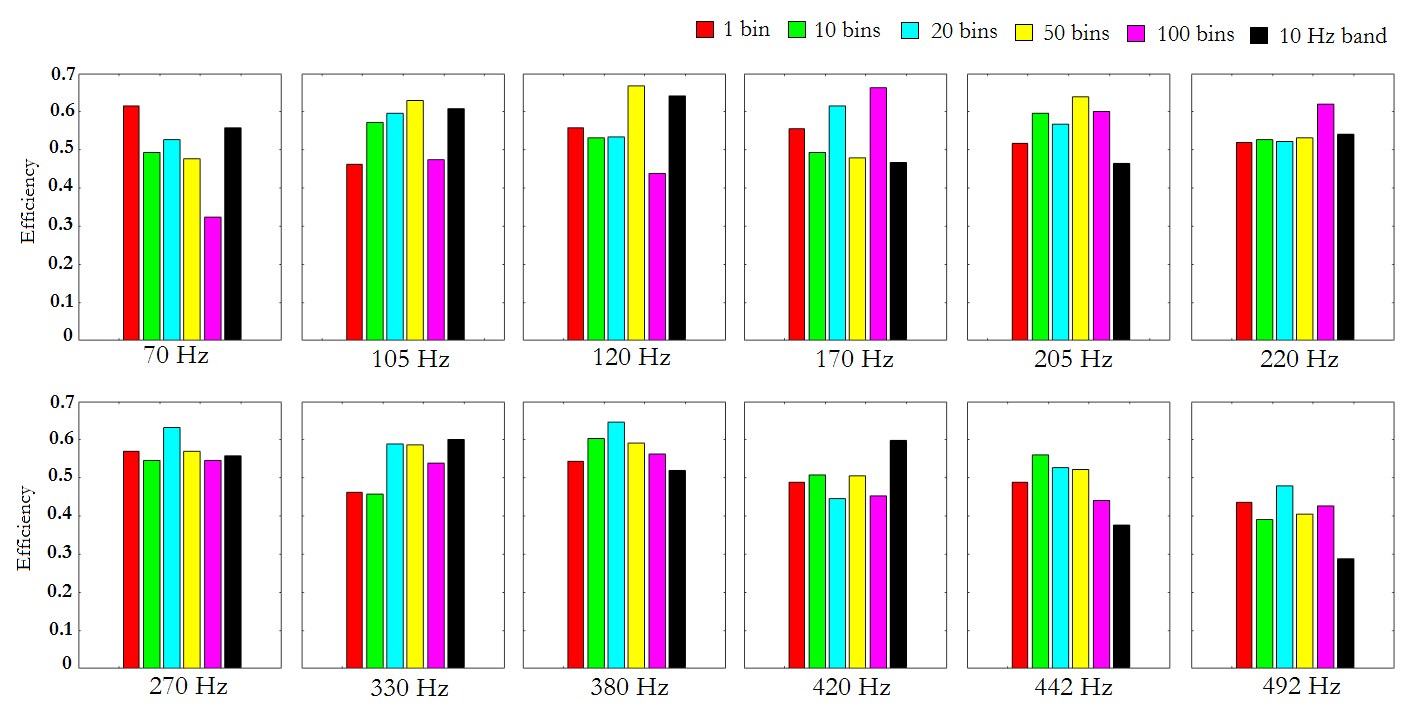}
\caption{Bar plots of the simulation done in Sec. \ref{subsec:lowdens}. The bars show the median efficiency computed for 10 signals with source frequency within respectively N$\delta\nu$ bins (N=1, 10, 20, 50, 100) and 10 Hz (40960$\delta\nu$ bins). The color code is: 10 signals in 1 (red) 10 (green) 20 (cyan) 50 (yellow) 100 (magenta) bins and 10 Hz (black). On the bottom of each box the central source frequency of the injected signals is shown. Signal amplitudes have been generated from a uniform distribution in the range $h_0=2 \times 10^{-25} - 1 \times 10^{-24}$. Those plots show that there is not an evident efficiency loss when signals are concentrated in smaller $\Delta f_0$.}
\label{fig:injTT}
\end{figure*}

\begin{figure*}
\includegraphics[width=13cm,height=8cm]{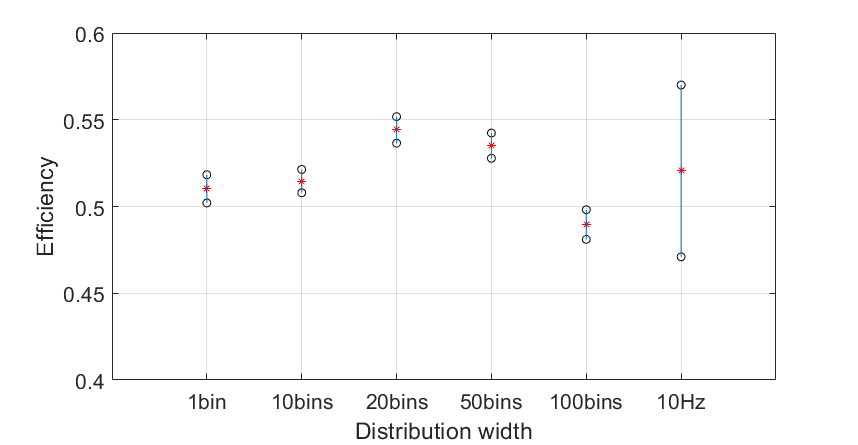}
\caption{Error plot of the simulation done in Sec. \ref{subsec:lowdens}, in the ``low-density" regime. Differently from Fig. \ref{fig:injTT}, efficiencies are combined together grouping for the different 12 frequency injections. Average efficiency $\eta$ (stars) and standard deviation (bars) are shown as function of $\Delta f_0.$}
\label{fig:injTT1}
\end{figure*}

Figure \ref{fig:injTT} shows bar plots of median values of the obtained efficiencies. They are grouped according to the different $\Delta f_0$ (represented by different colors), for the different injection frequencies. The efficiency values do not show any systematic decrease when going toward the smaller $\Delta f_0$, instead they seem more random. In Figure \ref{fig:injTT1} the bar plots are combined together and the average efficiencies (with respect to the injection frequency), together with standard deviations, are shown as a function of $\Delta f_0$. The highest standard deviation in the case of signals injected in 10 Hz is due to the fact that within a so large band it is more likely to find strong narrow disturbances that worsen the detection efficiency in that band. Also in this plot, there is no clear effect on efficiencies due to the clustering of source frequencies. We can conclude that in the low density regime there is not efficiency reduction and the procedure is able to recover all the signals.


\section{Simulation of signals clusters from boson clouds}
\label{sec:simBC}
In this section, we simulate situations that are expected in the most extreme cases of CWs emitted by boson clouds around galactic BHs. In the simulations proposed in \cite{ref:papaetc}, it has been suggested that the most extreme BHs - namely, those with spins close to the maximum allowed value - could potentially produce up to thousands of signals with amplitude above the O2 ULs at the detectors. Depending on galactic BH population and the boson mass, those signals could cluster into a frequency region having a width that spans from $\sim0.01\mathrm{Hz}$ to $\sim 1\mathrm{Hz}$. Based on these indications, we reproduced two scenarios, namely one with signals clustered into a [0.04 - 0.06] Hz band and another with signals in an order of magnitude larger $0.8$ Hz band. Also in these cases we repeated the simulations centering on different frequencies in the range [70 , 512] Hz.
Over 4,000 signals were generated with random parameters and amplitude in the range $\left[1\,- 20\right]\times10^{-25}$, following a power law distribution such to mimic the ratio of detectable signals below and above the UL as found in \cite{ref:papaetc}.

\begin{figure*}
\includegraphics[width=8cm,height=6cm]{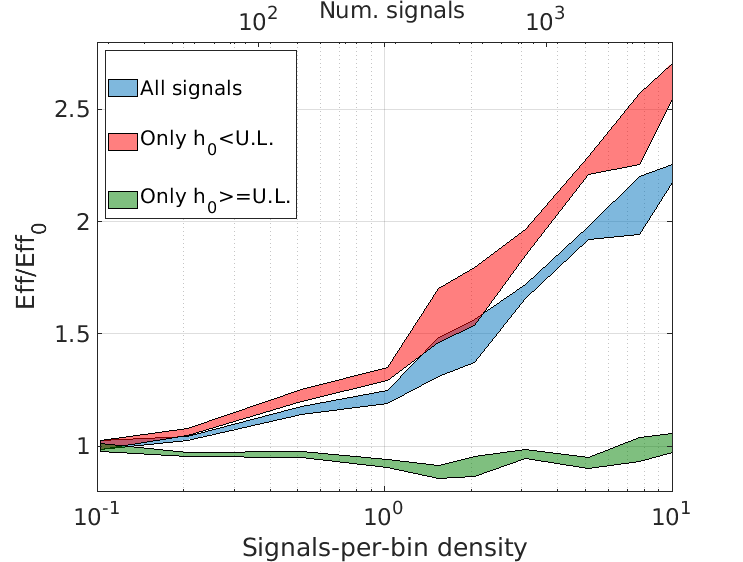}
\includegraphics[width=8cm,height=6cm]{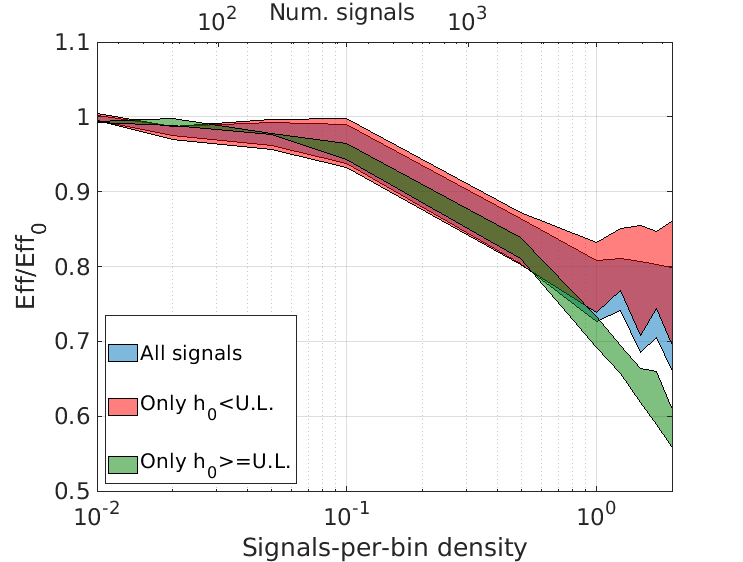}
\caption{ Injections in the ``high-density" regime. Mean values of the ratio between detection efficiencies in the case of signals clustered together ($\mathrm{Eff}$) and in the case of each of those signals taken alone ($\mathrm{Eff}_0$), as function of the signals-per-bin density (bottom x axis) and number of injected signals (top x axis). The signals are injected around 170 Hz, 240 Hz, 338 Hz, 380 Hz: colored areas span from minimum to maximum values obtained at the different frequencies. The blue area shows the overall efficiency ratio for the whole signal ensemble, whereas green and red areas refer to signals that are respectively above and below O2 ULs. Left plot shows the result when signals are clustered in a wide 0.06 Hz frequency range. Right plot shows the result when signals are clustered in a wide 0.8 Hz frequency range.}
\label{fig:eff_peaks_ratio}
\end{figure*}

In Figure \ref{fig:eff_peaks_ratio} the results of the simulation are shown. The plotted quantity is the ratio between the detection efficiency of signals in the clustered configuration and the efficiency of the same signals, when injected alone. This is done in order to compare efficiency variations at different frequencies and on signals with different amplitudes, which have different detection efficiencies. This efficiency ratio is plotted as function of the mean signal density. Colored areas span from minimum to maximum values of the mean ratio obtained from injections around different frequencies, in the range [100,400] Hz, in clean bands. The blue area represents the ratio for the whole ensemble of injected signals, whereas the green and red belts refer to the subsets of signals with $h_0$ respectively above and below O2 ULs.
Looking at the different regimes, we recognize qualitatively different behaviors. 

1) In the case of signals injected into a $\sim0.06$ Hz band (Figure \ref{fig:eff_peaks_ratio}, left plot), for signals with amplitude above O2 ULs, the efficiency loss ratio is of few percents at the most. After a minimum of $\approx 0.9$ reached between $\rho_{sig}\approx 1-2$, the ratio grows up to 1. On the other side, for signals with amplitude below O2 ULs the efficiency ratio increases for growing signal densities, which means their detection efficiency increases - i.e. we recover more pixels belonging to those signals. On the whole set of signals, the result is a global efficiency gain in the range $\left[1,2.2\right]$, the highest reached at the highest signal densities. An analogous reinforcement effect is used in \cite{ref:5nvect,ref:HC2F}, where the contributions of multiple subthreshold CWs are combined to enhance their detection chances in the context of targeted searches. Consequently, in this configuration there is no overall efficiency loss and, moreover, also weak signals have a growing probability to be recovered. Since also stronger signals do not suffer a significant efficiency loss, we conclude that both detection efficiency and evaluation of ULs are not affected.

2) The case of signals injected into a $\sim0.8$ Hz wide band (Figure \ref{fig:eff_peaks_ratio}, right plot) corresponds to the most extreme situation depicted in \cite{ref:papaetc}. In this case the behavior of the efficiency ratio changes significantly. First of all, the two distinct signal subsets have no opposite dynamics as they show both an efficiency loss as the signals density increases. While the efficiency ratio of signals with amplitude below O2 ULs stops decreasing at a signal density $\rho_{sig}\approx1$ (with a corresponding efficiency ratio of $\sim$ 0.8 $\pm$ 0.05), the efficiency ratio for the subset of signals with amplitude above the UL continues decreasing, reaching values around 0.6 for signal densities $\rho_{sig}\approx 2$. For the whole set of signals, values in the range 0.75 $\pm$ 0.05 are reached. Since the detection efficiency is directly related to the individual probability $p_{\lambda}$ to select a peak if a signal is present, and given that the overall search sensitivity is proportional to the square root of that probability \cite{ref:FH}, we conclude that in the most extreme - and probably very unrealistic - cases the search procedure could lose up to $\sim15\%$ of the ``optimal'' search sensitivity (i.e. that with no superposition of signals). As discussed in the following, a mean signal density of two is much larger than predicted in \cite{ref:papaetc}. 

Both the described dynamics have a clear explanation in the way the AR spectral estimator works. When the signal density is such to produce a wide excess power in frequency, the answer of the AR estimator depends on how wide the bump is with respect to the AR memory $\tau_f$ defined in Section \ref{sec:spectre}. The chosen value for AR memory is typically fixed for each frequency band. In the present case, in the frequency range [128-512] Hz, it is used $\tau_f=0.02$ Hz. In case 1), the signal cluster covers a wide $0.06$ Hz frequency range, which is the same order of $\tau_f$. Consequently, AR estimate is not fast enough to adapt to the increased noise (plus signal) level. When $\rho_{sig}\ge1$, every frequency bin has an expected occupation of 1 signal. After that, any added signal would superpose to signals already present in the same bin, thus increasing its power content. The result is that the signal power in all bins of interest increases with respect to the noise floor. If the AR estimation is not fast enough to adapt, it will be below the most part of bins power, so that also weaker signals get more detection chances. This explains the observed growing efficiency in case 1). On the other hand, in case 2) the signal cluster covers a wide $0.8$ Hz frequency range, which is one order of magnitude bigger than $\tau_f$. In this case, even if signals continue to accumulate in the same frequency bins at $\rho_{sig}\ge1$ and their power content increases, the AR estimation is able to adapt to the changed level. Thus, weak signals remain below the AR level, while strong signals are weakened with respect to the previous noise level.

We note two important considerations. First, the loss happens when we have actually high numbers of detectable signals (i.e., above the upper limit), while it affect much less the estimation of ULs. Second, we plot the efficiency loss as a function of the mean signal density. For instance, a density of two signals per bin, means that on average in \textit{all} bins of the injection band (e.g., 0.8 Hz in \ref{fig:eff_peaks_ratio}, right plot) we have such density. Looking at the results of \cite{ref:papaetc}, in one of their ``worst'' situations - see \cite{ref:papaetc} figure 40, left panel - where BH spins up to an unrealistically large value of 1 are considered, the signal density above the UL can reach values as large as 5 in a very few bins, while in the vast majority of bins is below 2 (and, actually, is zero in most bins). The average density is difficult to estimate, but it is quite likely well below 0.5. As a consequence, the sensitivity loss is likely of the order or smaller than 5$\%$, as the impact on the AR estimation would be smaller.  

\section{Discussion}
\label{sec:conc}
In this paper, mainly inspired by the claims of \cite{ref:papaetc}, we have studied in detail how the basic steps of the FH algorithm for CW searches could be affected by the presence of clusters of signals. Two regimes, of ``low-density" (with up to $\sim 0.1$ signals per frequency bin) and ``high-density" signals (with up to 10 signals per frequency bin), have been considered. In particular we have evaluated the impact of signal clusters in the estimation of the average spectrum, and on the construction of the PMs, finding them very robust. 
We have found that the procedure used to estimate the average spectrum, needed to normalize the FFTs and then to construct the PMs, is robust with respect to the presence of ensemble of signals and works even in the most extreme situations. The small signal peak amplitude loss is compensated by the fact that signals do have a different evolution in time (due to their position in the sky) and then through the proper Doppler correction each signal can be properly reconstructed. In addition to this, in the case of homogeneous high density of signals across a small frequency band, we have shown that when the signals are spread in a frequency interval smaller than $\sim0.1$ Hz, the overall efficiency increases thanks to the presence of signal clusters, as the probability to select a peak at a given time is enhanced. When the signals are spread on a much larger frequency range of $\sim0.8$ Hz, we find an efficiency loss of up to $\sim15\%$, when the mean signal density approaches a value of two, due to the impact the signals have in the AR estimation of the average power spectrum.
On the other hand, the ``worst'' cases considered in \cite{ref:papaetc} correspond to an mean signal density significantly smaller than two, so that the corresponding sensitivity loss is reduced to a few percent at the most.
\\ \textit{We conclude that the frequency-Hough procedure is robust with respect to the presence of signal ensembles, with negligible losses even in extreme cases.} 

The results reported here are relevant also in view of searches using the next generation detectors. The expected improved sensitivity of Einstein Telescope \cite{ref:futureET} with respect to current detectors, especially in the band 3-20 Hz, could make the issue of signal ensembles important. In particular, it could play a role not only for the case of emission from boson clouds around Kerr BHs, but especially for the early inspiral of NS binary systems, that would produce long duration signals and could be searched adapting techniques derived from standard CW searches.
Something similar is expected for the LISA detector \cite{ref:LISA}. Our result shows that the basic elements of FrequencyHough algorithm could be used to resolve most of these sources, exploiting their different sky position and the algorithm robustness with respect to the presence of signal ensembles.


\section*{Acknowledgement}
We thank the National Science Foundation for the international REU program, grants No. NSF PHY-1950830 and No. NSF PHY-1460803. \\
This research has made use of data or software obtained from the Gravitational Wave Open Science Center (gw-openscience.org), a service of LIGO Laboratory, the LIGO Scientific Collaboration, the Virgo Collaboration, and KAGRA. LIGO Laboratory and Advanced LIGO are funded by the United States National Science Foundation (NSF) as well as the Science and Technology Facilities Council (STFC) of the United Kingdom, the Max-Planck-Society (MPS), and the State of Niedersachsen/Germany for support of the construction of Advanced LIGO and construction and operation of the GEO600 detector. Additional support for Advanced LIGO was provided by the Australian Research Council. Virgo is funded, through the European Gravitational Observatory (EGO), by the French Centre National de Recherche Scientifique (CNRS), the Italian Istituto Nazionale di Fisica Nucleare (INFN) and the Dutch Nikhef, with contributions by institutions from Belgium, Germany, Greece, Hungary, Ireland, Japan, Monaco, Poland, Portugal, Spain. The construction and operation of KAGRA are funded by Ministry of Education, Culture, Sports, Science and Technology (MEXT), and Japan Society for the Promotion of Science (JSPS), National Research Foundation (NRF) and Ministry of Science and ICT (MSIT) in Korea, Academia Sinica (AS) and the Ministry of Science and Technology (MoST) in Taiwan.\\
We also thank the Amaldi Research Center, for the clusters hosted in Rome INFN, where we have stored the LIGO/Virgo data used in this research and run part of the present analysis. We thank the INFN-CNAF computing staff for the resources we have used in this analysis and for their constant support.\\
This project has received funding from the European Union's Horizon2020 research and innovation programme under the Marie Skłodowska-Curie grant agreement N. 754496.


\begin{thebibliography}{99}
\section{References}
\def\etal{{\it et al.}}

\bibitem{ref:lasky} P. D. Lasky, PASA 32, e034 (2015)

\bibitem{ref:CWreview} Piccinni, O. J. (2022). Galaxies, 10(3), 72.

\bibitem{ref:WP} The White paper of the LSC and Virgo collaboration (and Executive Summaries of searches)  \url{https://dcc.ligo.org/LIGO-T1700214/public}.

\bibitem{ref:GWTC1} LIGO Scientific collaboration, Virgo collaboration  \Journal{\PRX}{9}{031040}{2019}.

\bibitem{ref:GWTC2} LIGO Scientific collaboration, Virgo collaboration  \Journal{\PRX}{11}{021053}{2021}.

\bibitem{ref:GWTC2.1} LIGO Scientific collaboration, Virgo collaboration  arXiv:2108.01045, submitted to PRD.

\bibitem{ref:GWTC3} LIGO Scientific collaboration, Virgo collaboration, KAGRA Collaboration  arXiv:2111.03606, submitted to PRX.

\bibitem{ref:GW170817PM} LIGO Scientific collaboration, Virgo collaboration,  \Journal{\APJ}{875}{160}{2019}.

\bibitem{ref:LtransO3} Abbott, R., \etal \Journal{\APJ}{932}{133}{2022}.

\bibitem{ref:bosonTEOR} Brito, R., \etal \Journal{\PRD}{96}{064050}{2017}

\bibitem{ref:allskyO2} LIGO Scientific collaboration, Virgo collaboration,  \Journal{\PRD}{100}{024004}{2019}.

\bibitem{ref:allskyO3a} LIGO Scientific collaboration, Virgo collaboration, KAGRA Collaboration  \Journal{\PRD}{104}{082004}{2021}.

\bibitem{ref:allskyO3} LIGO Scientific collaboration, Virgo collaboration,  	arXiv:2201.00697 (2022), submitted to PRD. 

\bibitem{ref:CWmethods} Tenorio, R., Keitel, D., Sintes, A. M. (2021). Universe, 7(12), 474.

\bibitem{ref:FH}  P.Astone, A.Colla, S.DAntonio, S.Frasca, C.Palomba,\Journal{\PRD} {90} {042002} {2014}

\bibitem{IuriGPU} Iuri La Rosa, P. Astone, S. D' Antonio, \etal, {\em Universe} \textbf{7},218 (2021)

\bibitem{ref:papaetc} S. Zhu, M. Baryakhtar, M.A. Papa, et al. ,\Journal{\PRD} {102} {063020} {2020}.

\bibitem{BSD} O. J. Piccinni,  P. Astone, S. D' Antonio et al.,   \Journal{\CQG}{36:1}{015008}{2019}.

\bibitem{ref:pia_sfdb} P. Astone, S. Frasca and C. Palomba, Class Quantum Grav 22, S1197 (2005)

\bibitem{ref:UL} C. Palomba, S. D'Antonio, P. Astone, et al., \Journal{\PRL} {123} {171101} {2019}.

\bibitem{ref:nostroBOS} S. D'Antonio,  C. Palomba, P. Astone, et al. ,\Journal{\PRD} {98} {103017} {2018}

\bibitem{ref:bosonO3} LIGO Scientific collaboration, Virgo collaboration, KAGRA Collaboratio ,\Journal{\PRD} {105} {102001} {2022}

\bibitem{ref:BHpop1} A. Arvanitaki et al. ,\Journal{\PRD} {81} {123530} {2010}

\bibitem{ref:BHpop2} A. Arvanitaki et al. ,\Journal{\PRD} {83} {044026} {2011}

\bibitem{ref:futureCE} Reitze, D. \etal, (2019). arXiv preprint arXiv:1907.04833.

\bibitem{ref:futureET} M. Maggiore, C. Van Den Broek, N. Bartolo \etal {\em  arXiv:1912.02622v4 } {2020}.

\bibitem{ref:adele} A. La Rana, \emph{Eur. Phys. J. H} \textbf{47}, 3 (2022)

\bibitem{ref:futureL} S. Babak, J. Gair, A. Sesana et al, \Journal{\PRD} {95} {103012} {2017}, and references therein (this is the fifth science case).

\bibitem{ref:GWOSC} Abbott, R., \etal (2021).  SoftwareX, 13, 100658.

\bibitem{ref:5nvect} D’Onofrio, L., \etal, \Journal{\PRD} {105(6)} {063012} {2022}

\bibitem{ref:HC2F} Bennett, M. F., Melatos, \etal, \Journal{\APJ}{766(2)}{99}{2013} 


\bibitem{ref:LISA} Amaro-Seoane, Pau, et al., \CQG 29.12 (2012): 124016.

\end{thebibliography}
\end{document}